\documentclass[twocolumn,showpacs,preprintnumbers,amsmath,amssymb,superscriptaddress]{revtex4}
\usepackage{graphicx}
\usepackage{dcolumn}
\usepackage{bm}
\usepackage{color}
\usepackage{subfigure}
\usepackage{amsbsy}

\begin{document}

\title{Meissner effect in superconducting microtraps}
\author{D. Cano}
\author{B. Kasch}
\author{H. Hattermann}
\author{R. Kleiner}
\author{C. Zimmermann}
\author{D. Koelle}
\author{J. Fort\'{a}gh}

\affiliation{Physikalisches Institut, Eberhard-Karls-Universit\"at
T\"ubingen, CQ Center for Collective Quantum Phenomena and their
Applications, Auf der Morgenstelle 14, D-72076 T\"ubingen, Germany}

\begin{abstract}
We report on the realization and characterization of a magnetic
microtrap for ultra cold atoms near a straight superconducting Nb
wire with circular cross section. The trapped atoms are used to
probe the magnetic field outside the superconducting wire. The
Meissner effect shortens the distance between the trap and the wire,
reduces the radial magnetic-field gradients and lowers the trap
depth. Measurements of the trap position reveal a complete exclusion
of the magnetic field from the superconducting wire for temperatures
lower than 6K. As the temperature is further increased, the magnetic
field partially penetrates the superconducting wire; hence the
microtrap position is shifted towards the position expected for a
normal-conducting wire.
\end{abstract}

\maketitle

Microfabricated magnetic traps for cold atoms provide an intriguing
physical scenario in which solid-state and atomic physics converge.
Coherent control over the internal and external states of atomic
quantum gases has already been achieved by means of magnetic
potentials near microfabricated surfaces  \cite{Fortagh:07}. These
advances led also to a number of fundamental studies of atom-surface
interactions such as the spin decoherence of atoms near dielectric
bodies \cite{Henkel:99,Jones:03,Harber:03,Lin:04,Scheel:05}  and the
Casimir-Polder force between atoms and surfaces \cite{Obrecht:07}.
Superconductors are expected to play an essential role in this
emerging field of research because they can provide an extremely low
noise environment for trapped atoms \cite{Hohenester:07}. Increased
atomic coherence times at very short distances from superconducting
surfaces will allow the manipulation of atomic wave functions even
on the submicron scale. Also, it is likely that superconducting
microstructures will have important applications which combine
coherent atomic clouds with superconducting devices to form novel
hybrid quantum systems. They include atomic hyperfine transitions
coupled to local microwave sources made from Josephson junctions, or
even quantum computation with single dipolar molecules
\cite{Rabl:06} or Rydberg atoms that are coherently coupled by
superconducting electrodes \cite{Sorensen:04} .

Recent experiments demonstrated the feasibility of superconducting
microtraps \cite{Roux:08} and the trapping of atoms nearby a
persistent current loop \cite{Mukai:07}. The impact of the Meissner
effect on the microtrap parameters has been assessed so far only
theoretically \cite{Cano:08}. The experimental observation of this
fundamental property of superconductors requires short enough
distances between the atoms and the superconducting surface.
Understanding how the Meissner effect distorts magnetic potentials
is a prerequisite for a new experimental regime in which the
advantages of superconducting microstructures can be fully
exploited.

In this letter we report on the realization of a magnetic microtrap
for ultracold $^{87}$Rb atoms near a superconducting niobium wire
with circular cross section. By monitoring the position of the atom
cloud, we observe the Meissner effect to influence the magnetic
trap. Total field exclusion from the Nb wire is observed for
temperatures below 6 K. As the temperature of the Nb wire is
increased the magnetic field gradually penetrates into the wire.
Analytical expressions for describing the magnetic trap are deduced
by approximating the superconducting wire to a perfectly diamagnetic
cylinder.

\begin{figure}
\centerline{\scalebox{0.5}{\includegraphics{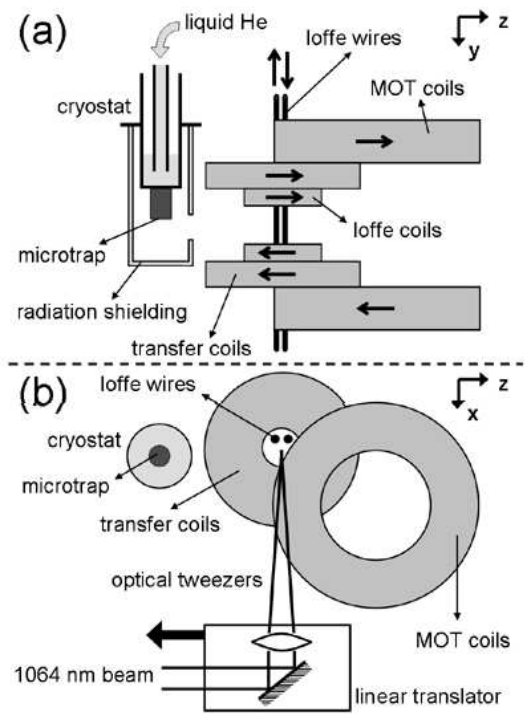}}}
\caption{Front view (a) and top view (b) of the experimental system
combining a standard room-temperature trap setup (to the right) and
a superconducting microtrap on a Helium flow cryostat (to the left).
Atoms are initially prepared in MOT and magnetic trap, and
subsequently transported to the superconducting microtrap by means
of optical tweezers. The room-temperature trap setup and the
cryostat are in a single vacuum chamber at a pressure of $10^{-11}$
mbar. The optical tweezers are translated over 44 mm with an
air-bearing translation stage outside the vacuum chamber. }
\label{figcamara}
\end{figure}

The experimental system integrates the techniques for producing
ultra cold atomic quantum gases with the techniques for cooling
solid bodies to cryogenic temperatures. A standard, room-temperature
trap setup for cooling atoms \cite{Fortagh:07} and a Helium flow
cryostat (ST-400 Janis) for operating superconducting
microstructures are installed next to each other in a single vacuum
chamber (Fig. \ref{figcamara}). The vacuum chamber is evacuated to
$10^{-11}$ mbar by an ion pump and a titanium-sublimation pump. The
transport of atoms from the room-temperature trap setup to the
cryogenically cooled superconducting microstructure is accomplished
by means of optical tweezers \cite{Gustavson:02}. Such an
arrangement is appropriate for trapping ultra cold atoms in
superconducting microstructures which have to be thermally isolated
from the environment and surrounded by a radiation shield.

\begin{figure}
\centerline{\scalebox{0.3}{\includegraphics{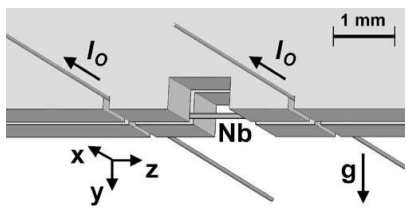}}}
\caption{Superconducting microtrap. The trap is generated below the
Nb wire by applying a current $I_{Nb}$ and a bias field $\bm B_B$ in
the positive $x$-direction. The Nb wire with circular cross section
of diameter 125 $\mu$m is clamped in a slit of a gold plated copper
holder which is attached to the cryostat. It can be cooled to
variable temperatures down to 4.2 K. A small cut in the copper
holder gives free access to the Nb wire at the central part of the
superconducting microtrap. A pair of kapton-insulated offset wires
along the $x$-direction, just above the Nb wire, are used for the
axial confinement of the trap. Gravity is along the $y$-direction.}
\label{figcopperpiece}
\end{figure}

The central piece of the experiment is a superconducting niobium
(Nb) wire with circular cross section of diameter 125 $\mu$m, as
shown in Fig. \ref{figcopperpiece}. The Nb wire is mechanically
clamped between two gold-plated copper plates which are firmly
attached to the Helium cryostat. This ensures mechanical stability
and good thermal contact between the wire and the cryostat. The Nb
wire is parallel to the $z$-axis. The microtrap is realized by
applying an electric current $I_{Nb}$ to the Nb wire and a
homogeneous bias field $\bm B_B$ along the $x$-direction
\cite{Fortagh:07}. Thus paramagnetic atoms can be radially confined
around a line parallel to the Nb wire, where $\bm B_B$ cancels the
circular magnetic field of $I_{Nb}$. The microtrap is closed in the
longitudinal direction by the magnetic field of two offset wires
driven with equal currents $I_0$ along the $x$-direction. The offset
wires are separated by 3mm. Since the Nb wire is electrically not
isolated from the Cu piece, the applied current $I_{Nb}$ will
entirely flow along the Nb wire only if this has no electrical
resistance. Therefore, no microtrap is generated in the
normal-conducting state. Transition to superconductivity is measured
at $T_c=9.2$ K with a four-point probe.

In order to load atoms into the superconducting microtrap, we follow
the experimental procedure as described below. Rubidium atoms
($^{87}$Rb) from a dispenser are trapped and cooled in a six-beam
MOT. After standard polarization gradient cooling and optical
pumping to the $|5S_{3/2},F=2,m_F=+2\rangle$ state the atoms are
trapped in the quadrupole magnetic trap generated by the MOT coils.
The atoms are then magnetically transferred via two transfer coils
into an Ioffe-Pritchard-type trap which is generated by two smaller
coils and a pair of wires parallel to the coil axis
\cite{Fortagh:07}, as shown in Fig. \ref{figcamara}. The strong
confinement of this trap with radial and longitudinal oscillation
frequencies of $\omega_r=2\pi\cdot272 \rm~s^{-1}$ and
$\omega_l=2\pi\cdot45\rm~s^{-1}$, respectively, allows for efficient
evaporative cooling. Radio-frequency evaporative cooling is applied
for 13 s to obtain a thermal cloud of $7 \cdot 10^5$ atoms at a
temperature of 2.5$\mu$K \cite{BEC}.

After evaporative cooling the atoms are transferred into an optical
dipole trap which is formed by focusing a 1064-nm laser with a
250-mm achromatic lens to a $1/e^2$ beam waist radius of 18$\mu$m.
The oscillation frequencies of the dipole trap are
$\omega_r=2\pi\cdot2100\rm~s^{-1}$ and
$\omega_l=2\pi\cdot30\rm~s^{-1}$. The transfer is accomplished by
ramping the laser light up to 500mW in 200 ms and then ramping down
the magnetic field in 20 ms. Next, the atomic cloud is transported
with the optical tweezers from the Ioffe-Pritchard-type trap over a
distance of 44mm to the cold surface of the cryostat. For this
purpose, the 250-mm lens is moved with an air-bearing linear
translation stage (Aerotech, ABL 1000), which is placed next to the
vacuum chamber (Fig. \ref{figcamara}). The air-bearing stage is
levitated with pressurized air and is driven by a brushless servo
motor that guarantees minimal vibration and that has an accuracy of
0.2 $\mu$m. Smooth transport is accomplished within 0.5 s using a
sinusoidal acceleration profile with a maximum acceleration of 1
m/s$^{2}$.

The atoms are loaded from the optical tweezers into the
superconducting microtrap at a distance of 500 $\mu$m from the Nb
wire. The loading is accomplished by ramping up the magnetic fields
of the microtrap within 100 ms, and subsequently ramping down the
laser power to zero in 0.5 s. The magnetic microtrap is generated by
$|\bm B_B|=0.64$ mT, $I_{Nb}=1.6$ A and $I_0=0.01$ A. A homogeneous
offset field $|\bm B_0|=0.1$ mT along the $z$-direction is
additionally applied to reduce Majorana losses \cite{Fortagh:07}.
Both $\bm B_B$ and $\bm B_0$ are generated by Helmoltz coils outside
the vacuum chamber. Trap frequencies are
$\omega_r=2\pi\cdot160\rm~s^{-1}$ and
$\omega_l=2\pi\cdot2\rm~s^{-1}$. The microtrap is typically loaded
with $4 \cdot 10^5$ atoms at 5$\mu$K.

\begin{figure}
\centerline{\scalebox{0.45}{\includegraphics{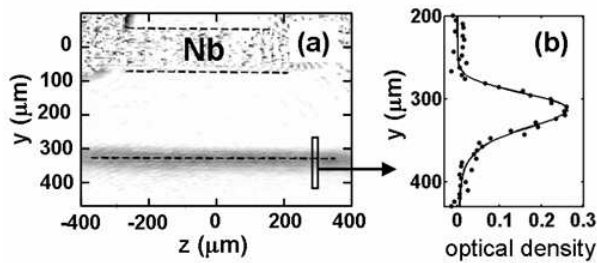}}}
\caption{(a) Absorption image of the atomic cloud at the central
region of the magnetic trap. The dashed lines represent the
longitudinal axis of the trap, and the Nb wire surface. (b) Optical
density of one of the transverse sections of the atomic cloud. The
best-fit curve $\rho (y)$ is also plotted.} \label{figatomos}
\end{figure}

As the atoms are forced into the minimum of the magnetic trap, they
can be used to probe the magnetic field profile near the
superconducting wire, and in this way to assess the Meissner effect.
The atoms are brought close to the Nb wire by reducing $I_{Nb}$
while keeping $\bm B_B$, $I_0$ and $\bm B_0$ constant. The
positioning of the cloud is accomplished within 0.5 s, which is
adiabatic with respect to the motion of atoms inside the trap. We
measure the position $y_0$ of the atom cloud with respect to the
center of the superconducting Nb wire for different wire currents
$I_{Nb}$. The data are derived from standard absorption images as
shown in Fig. \ref{figatomos}.

The atomic cloud in the microtrap is highly elongated due to the
weak longitudinal confinement. In order to determine the
longitudinal axis of the trap accurately, the absorption image of
the cloud is divided into transverse sections. For every transverse
section, the measured atom density is fitted by the theoretical atom
density $\rho (y)$ of a thermal cloud in a trapping potential
$U(y,x)$, as seen in Fig.\ref{figatomos}(b). The asymmetry of the
atom density profile is due to the decrease in magnetic gradient
with increasing $y$ and, to a lesser extent, due to gravity. These
two effects are included in the potential $U(y,x)$. The maxima of
$\rho (y)$ of all transversal sections of the atomic cloud are
fitted to a straight line, which is the longitudinal axis of the
microtrap. The value of $y_0$ is calculated as the distance between
the longitudinal axis of the trap and that of the Nb wire.

\begin{figure}
\centerline{\scalebox{0.35}{\includegraphics{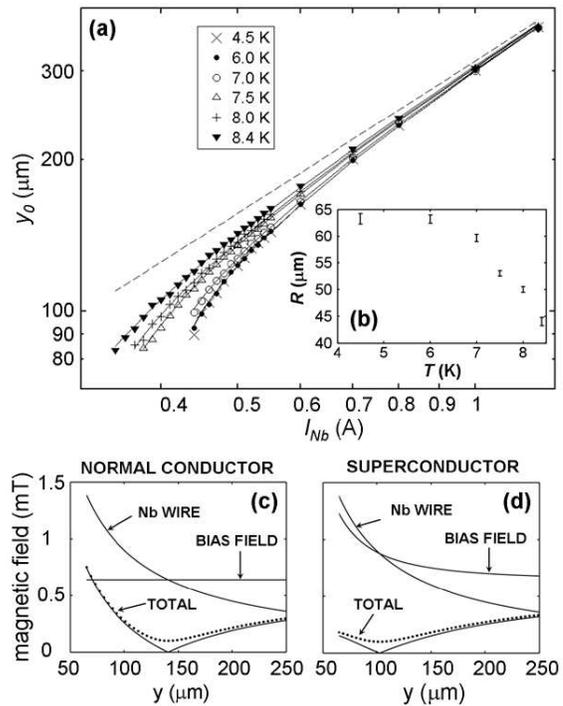}}}
\caption{ (a) Distance between the trap center and the wire center
as a function of the applied current $I_{Nb}$ for six different
temperatures. $|\bm B_B|=0.64$ mT, $I_{Nb}=1.6$ A and $B_0=10^{-4}$
T. The normal-conducting case is represented as a dashed curve. The
error bars, which result from the fitting procedure illustrated in
Fig. \ref{figatomos}, are smaller than the symbols and therefore not
represented. (b) Best-fit radius as a function of the Nb wire
temperature. The error bars represent the $95\%$-confidence
interval. (c) and (d) Modulus of the magnetic-field profiles
calculated for $I_{Nb}=0.45$A for the normal-conducting and the
superconducting cases, respectively. The comparison shows that the
Meissner effect strongly distorts the bias field while leaving the
field of $I_{Nb}$ unchanged. This results in a lower trap depth and
a shorter distance to the Nb wire. The realization of the microtrap
requires that the bias field is opposite in direction to $\bm
B_{Nb}$. The additional offset field $B_0=10^{-4}$T along the
$z$-direction changes the magnetic profile from linear to parabolic
(dotted curve).} \label{figtemperatura}
\end{figure}

Figure \ref{figtemperatura} (a) plots the trap position $y_0$ as a
function of $I_{Nb}$. Data are represented in logarithmic scale to
give more visibility to the points that are closer to the wire.
Measurements have been carried out for six different temperatures
$T$ of the Nb wire. The dependence of $y_0$ on $T$ is noticeable
only at short distances from the wire, where the impact of the
Meissner effect is stronger. For every temperature, the measured
data are fitted by a theoretical curve that assumes that the
superconducting Nb wire behaves like a diamagnetic cylinder of
effective radius $R$. For every temperature, the best-fit curve is
found by varying the effective radius $R$. Figure
\ref{figtemperatura}(b) represents the best-fit radius $R$ versus
$T$.

For comparison, the microtrap positions calculated for the
normal-conducting state are represented as a dashed curve. It is a
straight line, slightly distorted by gravity. Above $T_c$ of Nb,
$\bm B_B$ penetrates the conductive wire and remains homogeneous,
canceling the circular field of $I_{Nb}$ at a distance from the wire
center of $y_{0,NC} = ( \mu_0 I )/(2 \pi B_B)$. The magnetic-field
gradient in the radial directions around $y_{0,NC}$ is $a_{NC} =
B_B/y_{0,NC}$. The gravitational sag is a small quantity that can be
calculated as $\Delta_{NC}=(g m B_0)/(g_F \mu_B m_F a_{NC}^2)$
\cite{Fortagh:07}.

Formulas describing the magnetic trap for the superconducting case
are derived by approximating the Nb wire to a diamagnetic cylinder
of effective radius $R$. Because of the axial symmetry, the magnetic
field generated by $I_{Nb}$ outside the superconducting wire is the
same as in the normal case: $B_{Nb}=(\mu_0 I_{Nb})/(2 \pi y)$.
However, the bias field is strongly affected by the Meissner effect.
The bias field on the $y$-axis becomes $B_B$(1+$R^2$/$y^2$) $\bm
e_x$ \cite{explicacion}. In the superconducting case, the bias field
cancels the circular field of $I_{Nb}$ at a distance from the wire
center of
\begin{equation}
y_{0,SC} = \frac{\mu_0}{4 \pi} \frac{I_{Nb}}{B_B}+
\sqrt{\left(\frac{\mu_0}{4 \pi} \frac{I_{Nb}}{B_B}\right)^2-R^2}.
\label{y0SC}
\end{equation}
The magnetic-field gradient around $y_{0,SC}$ is
\begin{equation}
a_{SC} = \frac{B_B}{y_{0,SC}}
\left(1-\left(\frac{R}{y_{0,SC}}\right)^2\right). \label{aSC}
\end{equation}
The gravitational sag $\Delta_{SC}=(g m B_0)/(g_F \mu_B m_F
a_{SC}^2)$ is always smaller than 4$\mu$m and hardly affects the
best-fit radius $R$. Even so, the effect of gravity on the trap
position has been considered in the fitting procedure with the aim
of reducing the error of the obtained best-fit $R$. The measured
data are fitted by the function $y_{0,SC}+\Delta_{SC}$. For every
temperature, the root-mean-square of the difference between the
measured points and the best-fit curve is lower than 2$\mu$m, which
demonstrates that the magnetic field outside the Nb wire can be well
described by formulas relying on a diamagnetic cylinder.

The way in which the Meissner effect changes the trap parameters is
explained in Figs. \ref{figtemperatura} (c) and (d) by comparing the
magnetic profiles in the superconducting and in the
normal-conducting cases. The magnetic profile in the superconducting
case is calculated assuming that the effective radius equals the
real radius of the wire, $R=62.5\mu$m. The Meissner effect shortens
the distance between the trap and the wire, reduces the radial
magnetic-field gradients and lowers the trap depth. The fact that
the theoretical predictions for superconducting thin films reported
in Ref. \cite{Cano:08} follow a similar tendency suggests that this
does not depend on the particular geometry of the superconductor.

The longitudinal confinement, which is very weak at the trap center
because of the long distance between the offset wires, does not
alter the trap parameters described in the above equations. It is
also important to notice that the additional homogeneous offset
field $\bm B_0$ is not distorted by the Meissner effect because the
longitudinal demagnetizing factor of a long cylinder quickly tends
to zero as the strip length increases to infinity.

For temperatures lower than 6K, the best-fit radius $R$ is very
similar to the real radius of the wire. This is the expected value
at such low temperatures, when the wire is in the pure Meissner
state \cite{London:50}, i.e. when the magnetic flux penetrates the
superconducting wire only to the London depth, which is for Nb some
tens of nanometers \cite{Maxfield:65}. As the temperature is
increased, the microtrap positions are shifted towards the positions
expected for a normal-conducting wire. This is caused by an increase
in the amount of magnetic flux penetrating the wire, which is
manifested as a decrease of the effective radius $R$. The fact that
the increase in temperature does not affect the points that are far
from the wire demonstrates that the electric resistance remains
zero, and so the applied current $I_{Nb}$ flows entirely through the
Nb wire. The experimental data reveal a smooth transition from the
pure Meissner state to the normal state. For temperatures above 8.4
K, the atoms cannot be loaded into the microtrap because the
critical current of the Nb wire drops below 1.6 A, which is the
current $I_{Nb}$ required to load the microtrap.

Another set of 160 measurements has been taken at $T=4.5$ K with
different currents $I_{Nb}$ in the range of 0.3 to 1.8 A and
different bias fields in the range of 0.4 to 0.8 mT. By fitting the
function $y_{0,SC}+\Delta_{SC}$ to the overall data, we obtain an
effective radius of $(61.5 \pm 1.1) \mu$m, which corroborates the
presented results.

In conclusion, we demonstrated an experimental system that enables
studies at the interface of cold atoms and superconductors. We
measured the impact of the Meissner effect on the potential of a
magnetic microtrap near a superconducting wire. The position of the
atomic cloud reveals complete field exclusion from the
superconducting wire for temperatures below 6 K. For higher
temperatures, the microtrap parameters are sensitive to the
temperature of the superconducting wire. Even though transition from
the pure Meissner state to the normal state usually consists of
complex processes involving vortex formation and penetration-depth
increase \cite{Ketterson:99}, the trapping field outside the wire
can be well described by simple formulas relying on a diamagnetic
cylinder of effective, temperature-dependent radius $R$. The
Meissner effect will have important implications for experiments
with quantum gases near superconducting surfaces.

We thank Thomas Dahm for useful discussions. This work was supported
by the DFG (SFB TRR 21) and by the BMBF (NanoFutur 03X5506).


\begin{thebibliography}{10}

\bibitem{Fortagh:07} J. Fort\'{a}gh, and C. Zimmermann, Rev. Mod. Phys. \textbf{79}, 235 (2007).
\bibitem{Henkel:99} C. Henkel, S. P\"{o}tting, M. Wilkens, Applied Physics B \textbf{69}, 379 (1999).
\bibitem{Jones:03} M.~P.~A. Jones, C.~J. Vale, D. Sahagun, B.~V. Hall, and E.~A. Hinds, Phys. Rev. Lett. \textbf{91}, 080401 (2003).
\bibitem{Harber:03} D.~M. Harber, J.~M. McGuirk, J.~M. Obrecht, and E.~A. Cornell, J. Low Temp. Phys. \textbf{133}, 229 (2003).
\bibitem{Lin:04} Y.~J. Lin, I. Teper, C. Chin, and V. Vuletic, Phys. Rev. Lett. \textbf{92}, 050404 (2004).

\bibitem{Scheel:05} S. Scheel, P.~K. Rekdal, P.~L. Knight, and E.~A. Hinds, Phys. Rev. A \textbf{72}, 042901 (2005).
\bibitem{Obrecht:07} J.~M. Obrecht, R.~J. Wild, M. Antezza, L.~P. Pitaevskii, S. Stringari, and E.~A. Cornell, Phys. Rev. Lett \textbf{98}, 063201 (2007).
\bibitem{Hohenester:07} U. Hohenester, A. Eiguren, S. Scheel, and E.~A. Hinds, Phys. Rev. A \textbf{76}, 033618 (2007).
\bibitem{Rabl:06} P. Rabl, D. DeMille, J. M. Doyle, M. D. Lukin, R. J. Schoelkopf, and P. Zoller, Phys. Rev. Lett. \textbf{97}, 033003 (2006).
\bibitem{Sorensen:04} A.~S. Sorensen, C.~H. van der Wal, L.~I. Childress, and M.~D. Lukin, Phys. Rev. Lett. \textbf{92}, 063601 (2004).

\bibitem{Roux:08} C. Roux, A. Emmert, A. Lupascu, T. Nirrengarten, G. Nogues, M. Brune, J.~-M. Raimond, and S. Haroche, Europhysics Letters \textbf{81}, 56004 (2008).
\bibitem{Mukai:07} T. Mukai, C. Hufnagel, A. Kasper, T. Meno, A. Tsukada, K. Semba, and F. Shimizu, Phys. Rev. Lett. \textbf{98}, 260407 (2007).
\bibitem{Cano:08} D. Cano, B. Kasch, H. Hattermann, D. Koelle, R. Kleiner, C. Zimmermann, and J. Fort\'{a}gh, Phys. Rev. A \textbf{77}, 063408 (2008).
\bibitem{Gustavson:02} T. L. Gustavson, A. P. Chikkatur, A. E. Leanhardt, A. G\"{o}rlitz, S. Gupta, D. E. Pritchard, and W. Ketterle, Phys. Rev. Lett. \textbf{88}, 020401 (2001).

\bibitem{BEC} Bose-Einstein condensation is reached with $2 \cdot 10^5$ atoms by continuing evaporative cooling for another 5 s.
\bibitem{explicacion} This formula can be deduced from the surface electric current $\bm K = \bm M \times \bm n$ generated by the homogeneous magnetization $\bm M = - 2 \bm B_B$/$\mu_0$ in the dyamagnetic cylinder, where $\bm n$ is the unit vector perpendicular to the cylinder surface. A similar deduction for the sphere can be found in J.~D. Jackson \textit{Classical electrodynamics} (John Wiley $\&$ Sons, Inc., 1967), Chap. 5.
\bibitem{London:50} F. London, \textit{Superfluids} (Wiley, New York, 1950), Vol. I.
\bibitem{Maxfield:65} B.~W. Maxfield and W.~L. McLean, Phys. Rev. \textbf{139}, A1515 - A1522 (1965).

\bibitem{Ketterson:99} J.~B. Ketterson, and S.~N. Song, \textit{Superconductivity} (Cambridge University Press,1999).

\end{thebibliography}
\end{document}